\documentclass[smallextended]{svjour3}
\usepackage{natbib}
\usepackage{graphicx}
\usepackage{algorithm}        
\usepackage{algpseudocode}    
\usepackage{amssymb}
\usepackage{amsmath}
\usepackage{microtype}
\usepackage{graphicx}
\usepackage{tabularx}
\usepackage{caption}
\usepackage{pifont}
\usepackage{url}
\usepackage{tabularx}
\usepackage{makecell}
\usepackage{multirow}
\usepackage{ragged2e}
\usepackage{booktabs}
\DeclareMathOperator*{\argmax}{arg\,max}

\let\emptyset\varnothing
\newcommand{\In}{\textbf{in }}
\algnewcommand{\algorithmicgoto}{\textbf{Goto}}%
\algnewcommand{\Goto}[1]{\algorithmicgoto~\ref{#1}}%

\smartqed  
\usepackage{graphicx}
\begin{document}

\title{Disambiguating fine-grained place names from descriptions by clustering}
\subtitle{}

\titlerunning{Disambiguating fine-grained place names}        

\author{Hao Chen \and        
	Maria Vasardani \and
	Stephan Winter
}


\institute{Hao Chen \at \email{hchen@student.unimelb.edu.au} \and
           Maria Vasardani \at \email{maria.vasardani@unimelb.edu.au} \and
           Stephan Winter \at \email{winter@unimelb.edu.au} 
}

\date{Received: date / Accepted: date}

\maketitle

\begin{abstract}
Everyday place descriptions often contain place names of fine-grained features, such as buildings or businesses, that are more difficult to disambiguate than names referring to larger places, for example cities or natural geographic features. Fine-grained places are often significantly more frequent and more similar to each other, and disambiguation heuristics developed for larger places, such as those based on population or containment relationships, are often not applicable in these cases. In this research, we address the disambiguation of fine-grained place names from everyday place descriptions. For this purpose, we evaluate the performance of different existing clustering-based approaches, since clustering approaches require no more knowledge other than the locations of ambiguous place names. We consider not only approaches developed specifically for place name disambiguation, but also clustering algorithms developed for general data mining that could potentially be leveraged. We compare these methods with a novel algorithm, and show that the novel algorithm outperforms the other algorithms in terms of disambiguation precision and distance error over several tested datasets.

\end{abstract}

\section{Introduction}
\begin{figure}
	\centering
	\includegraphics[width=\textwidth]{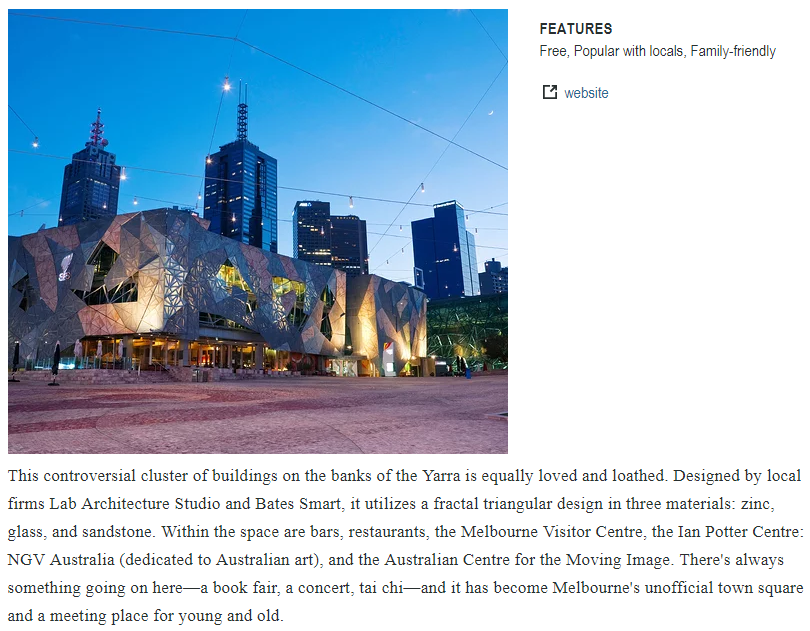}
	\caption{An example of a short description about Federation Square, a landmark in Melbourne, with several place names being mentioned (\textit{Source}: http://www.travelandleisure.com/travel-guide/melbourne/things-to-do/federation-square).}
	\label{fig:desc}
\end{figure}

Everyday place descriptions are a way of encoding and transmitting spatial knowledge about places between individuals \citep{vasardani2013descriptions,vasardani2013locating}. Also, the web provides a plethora of place descriptions such as news articles, social media texts, trip guides, and tourism articles \citep{kim2015harvesting}. An example of a place description from the web is shown in Figure~\ref{fig:desc}. For utilizing the expressed place-related knowledge in information systems the place names need to be identified and georeferenced (or located). A typical approach deploys place name gazetteers: directories of known names and their locations. Since many place names are ambiguous -- with multiple gazetteer entries -- the approach also includes a disambiguation process. The whole process consists of two steps: place name recognition (from text) and place name disambiguation, and is often called \textit{toponym resolution} \citep{leidner2008toponym}. This research focuses on the second challenge, i.e., place name disambiguation, with everyday place descriptions as the target document source.

Everyday place descriptions often contain place names of fine-grained features (e.g., names of streets, buildings and local points of interest). Most studies in the field of toponym resolution focus on larger geographic features such as populated places (e.g., cities or towns) or natural geographic features (e.g., rivers or mountains). For these features, disambiguation heuristics can leverage the size, population, or containment relationships of candidate places, possibly based on external knowledge bases (e.g., WordNet or Wikipedia). Such heuristics quickly fail when dealing with the fine-grained places in everyday place descriptions. Fine-grained places are often significantly more frequent and more similar to each other than those larger (natural or political) gazetteered places. Even disambiguation approaches based on machine-learning techniques are difficult to be applied for fine-grained places due to the lack of good-quality training data, as well as the challenge of locating previously-unseen place names. 

In this research we use map-based clustering approaches that have been developed for place name disambiguation. Map-based approaches should be relatively robust for fine-grained places as they only require knowledge of the locations of ambiguous candidate entries. However, it remains to be seen whether these algorithms are suitable for the task of this research. Some of them are defined for large geographic features and may not perform equally well on fine-grained places. Some algorithms are parameter-sensitive, and require manual input, and thus substantial pre-knowledge of the data. Therefore, we will also take a look at more generic clustering algorithms that exist in fields such as statistics, pattern recognition, and machine learning. In particular, we will compare existing clustering algorithms with a novel algorithm that is is designed to be robust, parameter- and granularity-independent. We will show that the new algorithm, despite being parameter-independent, achieves state-of-art disambiguation precision and minimum distance error for several tested datasets. 

The contributions of this paper are:

\begin{enumerate}
	\item a comparison of different clustering algorithms for disambiguating fine-grained place names extracted from everyday place descriptions;
	\item an in-depth analysis of algorithms from five categories (ad-hoc, density-based, hierarchical-based, partitioning relocation-based, and others) in terms of performance, reasons, and relative suitability of the task for each; and
	\item a new clustering algorithm which out-performs the other tested algorithms for the collected datasets.
\end{enumerate}

Accordingly, compared to existing algorithms, the advantages of the new algorithm are:

\begin{enumerate}
	\item it does not require manual input of parameter values and works well for data with different contexts, i.e., size of spatial coverage, distance between places, levels of granularity (parameter-independent).
	\item it achieves the highest average disambiguation precision and has overall minimal distance errors for the tested datasets, compared to other algorithms even with their best-performing parameter values. Note that these values are typically hard to determine without pre-knowledge of the data; and
	\item its performance is robust for descriptions with different contexts. Compared to other algorithms, it has low variation in both precision and distance error for different input data.
\end{enumerate}

The remainder of the paper is structured as follows: in Section \ref{sec:related} a review of relevant clustering algorithms is given. Section \ref{sec:new} proposes a new algorithm. Section \ref{sec:case} explains the input dataset as well as the experiment. Section \ref{sec:discussion} presents the obtained results as well as the corresponding discussions. Section \ref{sec:conclusion} concludes this paper.

\section{Related work} \label{sec:related}
In the following section, related work in disambiguating place names from text, as well as relevant clustering algorithms is introduced.

\subsection{Place name disambiguation}
Place name disambiguation, also known as \textit{toponym disambiguation}, is the task of disambiguating a place name with multiple corresponding gazetteer entries. For example, GeoNames\footnote{http://www.geonames.org/} lists 14 populated places `Melbourne' world-wide. Various approaches have been proposed in the past years mainly in the context of Geographic Information Retrieval (GIR), in order to georeference place names in text or geotagging whole documents. Typically, place name disambiguation is done by considering context place names, i.e., other place names occurred in the same document (discourse), and computing the likelihood of each of the candidate gazetteer entry to correspond this place name. The likelihood is computed as a score given some available knowledge of the context place names as well as the place name to be disambiguated, such as their locations or spatial containment relationships. For example, if `Melbourne' and `Florida' occur together in a document, then the place name `Melbourne' is more likely to be corresponding to the gazetteer entry `Melbourne, Florida, United States' rather than `Melbourne, Victoria, Australia'. There are also more recent language modeling approaches based on machine-learning techniques that not only consider context place names, but also other non-geographical words as well \citep[e.g.,][]{cheng2010you,roller2012supervised,wing2014hierarchical}. Many geotagging systems -- systems that determine the geo-focus for the entire document for geographic information retrieval purposes \citep[e.g.,][]{teitler2008newsstand,lieberman2007steward} -- heavily rely on place name recognition and disambiguation.

Depending on the knowledge used, disambiguation approaches can generally be classified into map-, knowledge-, and machine learning-based \citep{buscaldi2011approaches}. Map-based approaches rely mainly on the locations of the gazetteer entries of places names from a document, and use heuristics such as minimum point-wise distance, minimum convex hull, or closest to the centroid of all entries locations for disambiguation \citep[e.g.,][]{smith2001disambiguating,amitay2004web}. Previous studies that focus on disambiguating fine-grained places \citep[e.g.,][]{derungs2012resolving,moncla2014geocoding,palacio2015development}, are largely based on map-based approaches as well. Knowledge-based methods leverage external knowledge of places such as containment relationships, population, or prominence \citep[e.g.,][]{buscaldi2008conceptual,adelfio2013structured}. Machine learning-based approaches have the advantage of using non-geographical context words such as events, person names, or organization names to assist disambiguation, through creating models from training data representing the likelihood of seeing each of these context word associated with places \citep{smith2003bootstrapping,roberts2010toponym}. The selection of the disambiguation approach is usually task- and data source-dependent \citep{buscaldi2011approaches}, and it is also common that different approaches are used in hybrid manners.

\subsection{Relevant clustering algorithms} \label{sec:relevant}
Clustering is a division of data into meaningful groups of objects. A variety of algorithms exist, e.g., a review of clustering algorithms for data mining is given by \citet{berkhin2006survey}. In this section, we introduce clustering algorithms from two categories: ones that have been used for place name disambiguation before (including ad-hoc ones), as well as selected ones from the data mining community. These algorithms will be compared to the newly developed algorithm later in this paper. For the task of place name disambiguation, the input to these algorithms are the locations of all ambiguous candidate gazetteer entries of all place names from a document, in the form of a point cloud. 

\subsubsection{Clustering algorithms used for place name disambiguation} \label{sec:2.1}
\begin{figure}
	\centering
	\includegraphics[width=\textwidth]{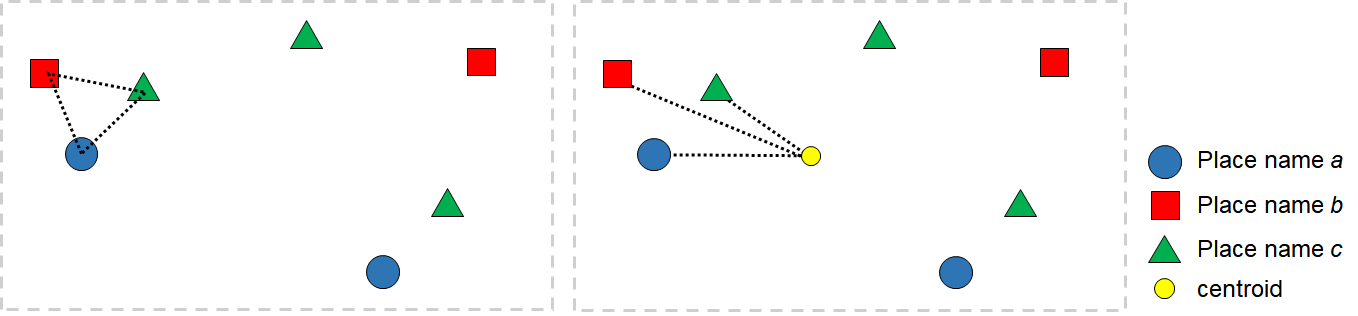}
	\caption{Clustering by overall minimum distance (left), and clustering by closeness to the centroid of all locations (right). Each symbol (other than the yellow one) represents the location of an ambiguous gazetteer entry of a place name.}
	\label{fig:clustering}
\end{figure}

The \textit{Overall minimum distance} heuristic aims at selecting gazetteer entries so that they are as geographically close to each other as possible. The closeness is typically measured either by the average location-wise distance, or area of the convex hull of these locations. An illustration of the algorithm is given in Figure~\ref{fig:clustering} (left): for each combination of ambiguous place name entries (one entry for each place name), create a cluster. Then, choose the minimum cluster representing the disambiguated locations, according to one of the measurement methods. This algorithm has been used in \citep{leidner2003grounding,amitay2004web,habib2012improving} and will generate only one cluster.
	
The \textit{centroid based} heuristic is explained in Figure~\ref{fig:clustering} (right). The algorithm first computes the geographic focus (centroid) of all ambiguous entry locations, and calculates the distance of each entry location to it. Then, two standard deviations of the calculated distances are used as a threshold to exclude entry locations that are too far away from the centroid. Next, the centroid is recalculated based on the remaining entry locations. Finally, for each place name, select the entry that is closest to the centroid for disambiguation. The algorithm is used in \citep{smith2001disambiguating,buscaldi2008map} and will also derive only one cluster. 

The \textit{Minimum distance to unambiguous referents} heuristic consists of two-steps. It first identifies unambiguous place names, i.e., place names with only one gazetteer entry, or place names that can be easily disambiguated based on some heuristics (e.g., when the method is used in conjunction with knowledge-based methods). Then, use a scoring function for the disambiguation of the remaining ambiguous entries, such as based on average minimum distance to those unambiguous entry locations, or weighed average distance considering times of occurrence in document or textual distance. The method appears in \citep{smith2001disambiguating,buscaldi2010grounding} and again will generate one cluster.

The DBSCAN algorithm (Density Based Spatial Clustering of Applications with Noise) is a density-based method that relies on two parameters: the neighborhood distance threshold $\varepsilon$, and the minimum number of points to form a cluster \emph{MinPts}. There is no straightforward way to fit the parameters without pre-knowledge of the data. Moncla \emph{et al.} use DBSCAN for the purpose of place name disambiguation \citep{moncla2014geocoding}, and the parameters in their case were empirically adjusted, since the authors have good understanding of the spatial coverages of the input data as hiking itineraries. A heuristic is proposed to estimate the value of parameters based on \textit{k-dist graph} (a line plot representing the distances to the \textit{k}-st nearest neighbor of each point) in the paper of DBSCAN \citep{ester1996density}. However, it is not trivial to detect the threshold, which requires a selection of value \textit{k} as well as knowledge of the percentage of noise within the data.

\subsubsection{General clustering algorithms for data mining} \label{sec:2.2}
This section introduces clustering algorithms from four groups: density-based, hierarchical-based, partitioning relocation-based, and uncategorized ones.

Using DBSCAN requires a-priori knowledge of the input data to determine the parameters. Some data, such as everyday descriptions in this research, have potentially various conversational contexts, and thus distances between the places mentioned. The algorithm OPTICS (Ordering Points To Identify the Clustering Structure) \citep{ankerst1999optics} address the problem by building an augmented ordering of data which is consistent with DBSCAN, but covers a spectrum of all different $\varepsilon' \leq \varepsilon$. The OUTCLUST algorithm exploits local density to find clusters that are mostly deviating from the overall population (clustering by exceptions) \citep{angiulli2006clustering} given \textit{k}, the number of nearest neighbors for computing local densities, as well as \textit{f}, a frequency threshold, for detecting outliers. 

Hierarchical clustering algorithms typically build cluster hierarchies and flexibly partition data at different granularity levels. The main disadvantage is the vagueness of when to terminate the iterative process of merging or dividing subclusters. CURE (Clustering Using REpresentatives) \citep{Guha1998cure} samples an input dataset and uses an agglomeration process to produce the requested number of clusters. CHAMELEON \citep{karypis1999chameleon} leverages dynamic modelling method for cluster aggregation considering k-nearest neighbor connectivity graph. HDBSCAN \citep{campello2013density} extends DBSCAN based on excluding border-points from the clusters and follows the definition of density-levels.

Partitioning relocation clustering divides data into several subsets, and certain greedy heuristics are then used for iterative optimization. The KMeans algorithm \citep{hartigan1979algorithm} divides the data into \textit{k} clusters through some random initial samples as well as an iterative process to update the centroids of the clusters until convergence. A Gaussian Mixture Model (GMM) \citep{celeux1992classification} attempts to find a mixture of probability distributions that best model the input dataset through methods such as the Expectation-Maximization (EM) algorithm. KMeans is often regarded as a special case of GMM.

There are other algorithms that do not belong to the previous three categories. The SNN (Shared Nearest Neighbours) algorithm \citep{ertoz2003finding} blends a density based approach by first constructing a linkage matrix representing the similarity, e.g., distance, among shared nearest neighbors based on $k$-nearest neighbors (KNN). The remaining part of the algorithm is similar to DBSCAN. Spectral clustering relies on the eigenvalues of the similarity matrix (e.g., KNN) of the data and performs partition of the data into the required number of clusters. Compared to KMeans, spectral clustering cares about connectivity instead of compactness (e.g., geometrical proximity). Kohen's Self Organizing Maps (SOM) \citep{kohonen1998self} is an artificial neural network-based clustering technique applying competitive learning using a grid of neurons. It is able to perform dimensionality reduction and map high-dimensional data to (typically) two-dimensional representation. 

\section{A new robust, parameter-independent algorithm} \label{sec:new}
The task of this research is the following: Given a place description $D$ with $i$ gazetteered place names extracted, $\{p_{1},p_{2},\ldots,p_{i}\}$, each name has a set of (one or more) corresponding gazetteer entries $\{p_{i}^1,p_{i}^2,\ldots,p_{i}^j\}$ that it can be matched to. In order to disambiguate each place name and link it to the entry that it is actually referring to (e.g., $p_{i}$ to $p_{i}^2$), clustering algorithms can be used to either minimize the geographic distances between the disambiguated entries according to some objective function (e.g., minimal average pairwise distance), or to derive high-density clusters that are likely to represent the geographic extents where the original descriptions are embedded. The input to such a clustering algorithm is a 2-dimensional point cloud with the locations of all ambiguous entries $p_{m}^n, m \in (1, i)$. 

The task is to select clusters by these objectives rather than to classify input data into several clusters. Such clusters will then be used for disambiguation, since they are expected to capture the true entries that the place names actually refer to. Points not captured by these clusters will be regard as noise. Therefore, certain clustering algorithms seem more suitable for this task than others, e.g., DBSCAN over KMeans. Furthermore, algorithms that are not parameter-sensitive or require no parameter are preferable, as place descriptions may have various spatial coverages, distance between places, and levels of granularity, thus no pre-knowledge can be assumed. In this section, we propose a novel density-based clustering algorithm \textit{DensityK}. The algorithm is robust, parameter-independent, and consists of three steps.

\subsection{Step one: computing point-wise distance matrix} \label{3.2}
In the first step, the algorithm computes all point-wise distances of an input point cloud, and the time complexity is $O(n^2)$ ($n$ is the number of input points). The time complexity can be reduced to $O((n^2-n)/2)$ with a distance dictionary to avoid re-computation (but needs $O(n^2)$ memory). The worst case time complexity is equal to DBSCAN, both without any indexing mechanism for neighborhood queries. In practice, DBSCAN is expected to be faster since it requires a defined distance threshold $\varepsilon$ and only considers point-wise distances below or equal to the value. With an index, e.g., R-Tree, the computation time can be reduced. $O(n^2)$ is also the worst case time complexity for algorithms that require computing neighborhood distances, e.g., OUTCLUST, SNN, and HDBSCAN. Still, a distance upper-bound value can be enforced for DensityK as an optional parameter to facilitate proceeding time, with an indexing approach similar to DBSCAN.

\subsection{Step two: deriving cluster distance}
In the second step, DensityK analyzes the computed point-wise distances, and derives a cluster distance automatically. The cluster distance is similar to the parameter $\varepsilon$ in DBSCAN, and will be used in the next step for generating clusters. 

First, a \textit{DensityK function} is computed given the point-wise distances in the first step, as shown in Function~\ref{eq:dk}. $K(d)$ represents the average point density for points within a given distance interval $(d-\Delta d, d]$ for all points in an annular region. The reason to apply annular search region for computing point density instead of circular region (i.e., $\Delta d = d$) is because we found the former one leads to better clustering results. A comparison of applying the two search regions is given later in this section. In Function~\ref{eq:dk}, the expression $count(p\in region(p_{i}, {(d-\Delta d, d]}))$ represents the number of points that are at a distance between $d-\Delta d$ and $d$ (including $d$) from point $p_{i}$. If there is no point within all the search regions for all points for a distance interval $(d_{j}-\Delta d, d_{j}]$, skip to the next interval ($(d_{j}, d_{j}+\Delta d]$). Thus, $K(d)$ is aways positive. The denominator of the left side of the function is the area of the annular region. $\Delta d$ is for discretizing the function and is set to 100$m$ in this research. The resulting cluster distance threshold will be the integer multiple of $\Delta d$. We will demonstrate below in this section that the clustering result is little sensitive to the value of $\Delta d$. 

\begin{equation}
\label{eq:dk} 
K(d) = \frac{1}{\pi d^{2} - \pi (d-\Delta d)^{2}} \times \frac{1}{n}\sum_{i=1}^{n} count(p\in region(p_{i}, {(d-\Delta d, d])})
\end{equation}

The approach is inspired by Ripley's $K$ function \citep{ripley1976second} which was originally designed to assess the degree of departure of a point set from complete spatial randomness, ranging from spatial homogeneity to a clustered pattern. Ripley's K function cannot be used to derive clusters nor cluster distances, yet the idea of detecting point density accordingly to distance threshold  meets our interest. The goal of this research is to derive a cluster distance threshold which leads to clusters with significantly large point densities. DensityK is a new algorithm with a different purpose than Ripley's $K$ function, but Ripley's $K$ function can be regarded as a cumulative version of the DensityK function. If the point-wise distances from the last step are sorted, the time complexity of computing DensityK function is $O(n)$ as it makes at most \textit{n} comparisons regarding different values of $d$. 

\begin{figure}[h]
	\centering
	\includegraphics[width=\textwidth]{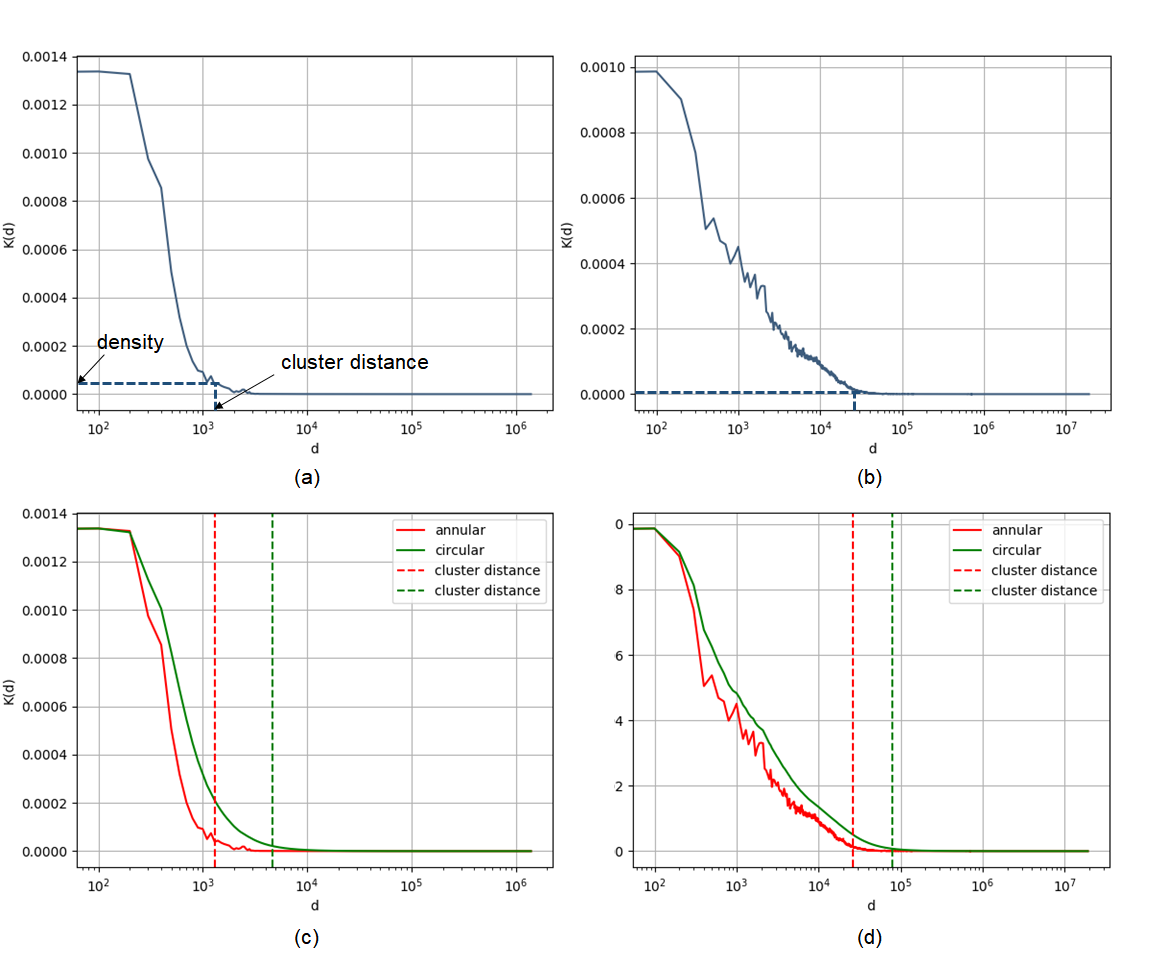}
	\caption{Two example DensityK functions from different input data with cluster distance highlighted (a, b), and comparisons of DensityK functions generated based on annular and circular search regions for the same data as in (a) and (b) respectively (c, d).}
	\label{fig:densityk}
\end{figure}

The function is able to show values of $d$ with significantly large point densities. Two illustrative examples are given in Figure~\ref{fig:densityk} (a) and (b) with different input data. For each of the two sample functions, $K(d)$ starts at a non-zero value for the first $d$: 100$m$ (the value of $\Delta d$), which means there are some points that are within 100$m$ from other points in the input point cloud. As $d$ grows, the value of $K(d)$ continues to decrease. For different input data, it is also possible that $K(d)$ starts from a low value, and then increases until a maximum value is reached, after which it starts to decrease again. 

Next, the mean $\mu$ and standard deviation $\sigma$ of all $K(d)$ values (a finite set since the function is discretized by $\Delta d$) are calculated. Then, the 2$\sigma$ rule is applied, and the minimum value of $d$ is selected as the cluster distance, that is $d > d_0$, $d_0 = \argmax_d K(d)$ and $K(d) = \mu + 2\sigma$. The derived cluster distances are also shown in Figure~\ref{fig:densityk} (a), (b). Intuitively, the cluster distance is the value of $d$ at the `valley' of a DensityK - a visually identifiable (at least roughly) x-value where the decrease pace of $K(d)$ value dramatically changes, leading to values close to zero. It is found that the resulting cluster distances always sit somewhere at the `valley' of the functions (in terms of $K(d)$ values) for different input data, and the derived clusters afterwards match quite well to the actual \textit{spatial contexts} (spatial extents where the descriptions are embedded). 

A comparison of annular and circular (replacing all $\Delta d$ by $d$ in Function~\ref{eq:dk}) search regions is shown in Figure~\ref{fig:densityk} (c) and (d), with the same input data as in (a) and (b) respectively. When tested on sample data, it is found that when applying annular regions, the derived clusters are always more constrained (as the computed cluster distances are smaller) and closer to the actual spatial contexts than those derived from circular regions. Such more constrained clusters are preferred as they are more likely to exclude ambiguous entries. It is found that they lead to higher disambiguation precision on the tested data as well. This phenomenon is most likely because when applying annular regions, the DensityK functions are more sensitive to the change of local density. In comparison, applying circular regions results in smoother density functions and possibly much larger cluster distances derived.

\begin{figure}[h]
	\centering
	\includegraphics[width=0.6\textwidth]{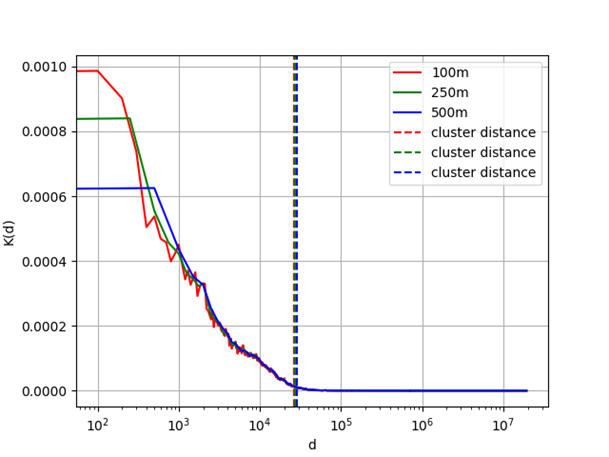}
	\caption{DensityK function generated with four different $\Delta d$ intervals for the same input point cloud: 100, 250, 500 (meters).}
	\label{fig:interval}
\end{figure}

DensityK function is little sensitive to the value of $\Delta d$. As shown in Figure~\ref{fig:interval}, the DensityK function plots generated for the same input data with three different $\Delta d$ values 100, 250, and 500$m$ are similar, and the cluster distances derived are the same. $\Delta d$ should be set to a constant, small number (e.g., the values in Figure~\ref{fig:interval}) for all input data, just for the purpose of discretization. Such a small number works well for various input data, even those with large cluster distances. Note that there is no single-optimal cluster distance for disambiguation. For example, different cluster distances from $2500m$ and $3500m$ may lead to the same disambiguation result for a given input; however, a cluster distance with value $25000m$ for the same input may increase or reduce the disambiguation precision, depending on the distances between the actual locations of the place names. 

Algorithm~\ref{alg:1} explains the whole procedure of this step, with sorted point-wise distances from the last step as input. The first part of the algorithm computes $K(d)$ for different $d$ values, and stores tuples of $(d, K(d))$ in the list variable $\textit{KFunction}$. Then, the cluster distance is derived given $\textit{KFunction}$. 

\begin{algorithm}[h]
	\caption{Computing cluster distance threshold.}
	\label{alg:1} 
	\textbf{Input:} $PointWiseDistances$: a sorted list of distance floats in meters \\
	\textbf{Output:} $ClusterDistance$: a float in meters
	\begin{algorithmic}[1]
		\State $\textit{KFunction} :=$ an empty list of 2-element tuples
		\State $\textit{MaxDistance} :=$ maxValue($\textit{PointWiseDistances}$)
		\State $\textit{NumberOfDistances} :=$ length($\textit{PointWiseDistances}$)
		\For {$d$ \In iterate($0$, $\textit{MaxDistance}$, $\Delta d$)} \Comment{loop of (min, max, interval)}
			\State $\textit{PointCountInRadius} := 0$
			\For {$\textit{distance}$ \In $\textit{PointWiseDistances}$}
				\If {$\textit{distance} \leq d$}
					\State $\textit{PointCountInRadius}$ += 1
				\EndIf
			\EndFor
			\If {$\textit{PointCountInRadius} > 0$}
			\State $\textit{Area} := \pi (d^2 - (d-\Delta d)^2)$
			\State $\textit{Density} := \dfrac{\textit{PointCountInRadius}}{\textit{Area} \times \textit{NumberOfDistances}}$
				\State $\textit{KFunction} \gets (d, \textit{Density})$ \Comment{Function~\ref{eq:dk}}
			\EndIf
		\EndFor
		\\
		\State $\textit{Densities} :=$ getDensities($\textit{KFunction}$)
		\State $\textit{Mean}, \textit{StandardDeviation} :=$ getMeanAndStd($\textit{Densities}$)
		\State $\textit{ThresholdDensity} := \textit{Mean} + 2 \times \textit{StandardDeviation}$
		\State $\textit{ClusterDistance} :=$ getCorrespondingDistance($\textit{ThresholdDensity}, \textit{KFunction}$)
		\State \Return $\textit{ClusterDistance}$
	\end{algorithmic}
\end{algorithm}

\subsection{Step three: deriving clusters and disambiguation}
The procedure of deriving clusters is similar to DBSCAN. Points that are within the cluster distance threshold are merged into the same cluster. The last step is to assign each place name with a location for disambiguation. To do so, the derived clusters are ranked by their contained number of points in descending order. Then, for each place name, choose the entry that first appears in one of the cluster according to the ranking, and the first cluster an entry appears is called a \textit{top-cluster} for this place name. For example, if an entry of a place name appears in the cluster with the largest number of points, the entry will be selected for disambiguation. If no corresponding entry of the place name is found in the first cluster, then the next cluster is chosen, until one entry is found. Thus, the worst case time complexity of this step is $O(nm)$ ($m$ is the number of clusters derived). In practice, as most places names are expected to be located within the first cluster, the time complexity is close to $O(n)$. The reason that we consider multiple clusters derived instead of only the first cluster is because it is possible that the input place names are from multiple spatial foci, i.e., the locations of some of the named places are relatively far away. In such cases, these isolated place names will be missed by the first cluster thus cannot be disambiguated correctly. The complete disambiguation procedure of this step is given in Algorithm~\ref{alg:2}.

\begin{algorithm}
	\caption{Disambiguation using the derived clusters.}
	\label{alg:2} 
	\textbf{Input:} $\textit{Clusters}$, $\textit{PlaceNamesAndEntries}$ as an list of 2-element tuples $\{(p_{i}, entry_{ij}),\ldots\}$ \\
	\textbf{Output:} $\textit{DisambiguatedPlaceNames}$
	\begin{algorithmic}[1]
		\State $\textit{DisambiguatedPlaceNames} := \emptyset$
		\State $\textit{RankedClusters} :=$ rankDescendent($\textit{Clusters}$)		
		\For {$\textit{Place}$ \In getPlaces($\textit{PlaceNamesAndEntries}$)} \label{marker}	 	
			\For {$\textit{Cluster}$ \In $\textit{RankedClusters}$}					
				\For {$\textit{Entry}$ \In getCorrespondingEntries($\textit{Entry}$, $\textit{PlaceNamesAndEntries}$)}	
					\If {$\textit{Entry}$ \In $\textit{Cluster}$}							
						\State $\textit{DisambiguatedPlaceNames} \gets (\textit{Place}, \textit{Entry})$
						\State \Goto{marker}
					\EndIf													
				\EndFor
			\EndFor	
		\EndFor		
		\State \Return $\textit{DisambiguatedPlaceNames}$
	\end{algorithmic}
\end{algorithm}

\section{Experiment on comparison of the clustering algorithms} \label{sec:case}
This section describes the input datasets, preprocessing procedure, used gazetteer and parser, and the final input to the algorithm to be tested. Then, the experiment settings in terms of algorithms and values used for their parameters are introduced.

\subsection{Dataset and preprocessing}
Two sets of place descriptions are used in the experiment. The first one contains 42 descriptions submitted by graduate students about the University of Melbourne campus, which are relatively focused in spatial extent \citep{vasardani2013descriptions}. The second one was harvested from web texts (e.g., Wikipedia, tourist sites, and blogs) about places around and inside Melbourne, Australia \citep{kim2015harvesting}. The two datasets cover more than 1000 distinct gazetteered places. Two example descriptions from the two datasets are shown below respectively, with gazetteered place names highlighted:

\begin{quotation}
	``... If you go into the \textbf{Old Quad}, you will reach a square courtyard and at the back of the courtyard. You can either turn left to go to the \textbf{Arts Faculty Building}, or turn right into the \textbf{John Medley Building} and \textbf{Wilson Hall} [...] If you continue walk along the road on the right side where you're facing \textbf{Union House}, you can see the \textbf{Beaurepaire} and Swimming Pool. There will also be a sport tracks and the \textbf{University Oval} behind it ...''
\end{quotation}

\begin{quotation}
	``... \textbf{St Margaret's School} is an independent, non-denominational day school with a co-educational primary school to Year 4 and for girls from Year 5 to Year 12. The school is located in \textbf{Berwick}, a suburb of \textbf{Melbourne} [...] In 2006, St Margaret's School Council announced its decision to establish a  brother school for St Margaret's. This school opened in 2009 named \textbf{Berwick Grammar School}, currently catering for boys in Years 5 to 12, with one year level being added each year ... ''
\end{quotation}

\begin{figure}[!h]
	\centering
	\includegraphics[width=\textwidth]{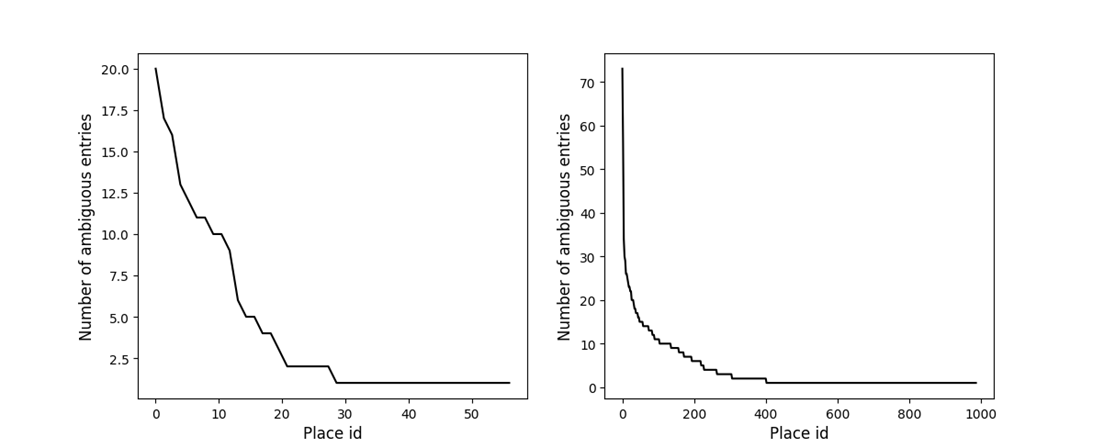}
	\caption{Numbers of ambiguous gazetteer entries of places names from the two datasets, campus (left) and Melbourne (right).}
	\label{fig:ambiguity}
\end{figure}

Place name recognition is outside the scope of this research, and we used a previously-developed parser to extract place names from each of the descriptions \citep{liu2013automatic}. Then, three gazetteers were used in conjunction for retrieving (ambiguous) entries for the extracted names, aiming for completeness: OpenStreetMap Nominatim geocoder \footnote{https://nominatim.openstreetmap.org/}, GoogleV3 geocoder \footnote{https://developers.google.com/maps/documentation/geocoding/intro}, and GeoNames \footnote{http://www.geonames.org/}. For example, the name \textit{St Margaret's School} has a total of 11 corresponding entries from the three gazetteers. The retrieved entries from the three sources were then synthesized, and duplicated entries referring to the same places were removed. The numbers of ambiguous gazetteer entries retrieved are shown in Figure~\ref{fig:ambiguity}, representing the ambiguities of these place names. 

Next, the extracted place names are manually linked to their corresponding gazetteer entries to create the groundtruth data for evaluation. For each description document, the input to the algorithms to be tested in the experiment below is the locations of all ambiguous gazetteer entries of place names extracted from the document, as a point cloud. An illustrative example is provided below in Figure~\ref{fig:input} based on a document from the campus dataset. The ground truth locations of these place names (the locations of their corresponding gazetteer entries), which are inside or near the University of Melbourne campus, are highlighted by red color in the bottom-right corner. For the algorithms to be tested below, each place name is considered as a successful disambiguation if it is correctly linked to its corresponding gazetteer entry.

\begin{figure}[h]
	\centering
	\includegraphics[width=0.7\textwidth]{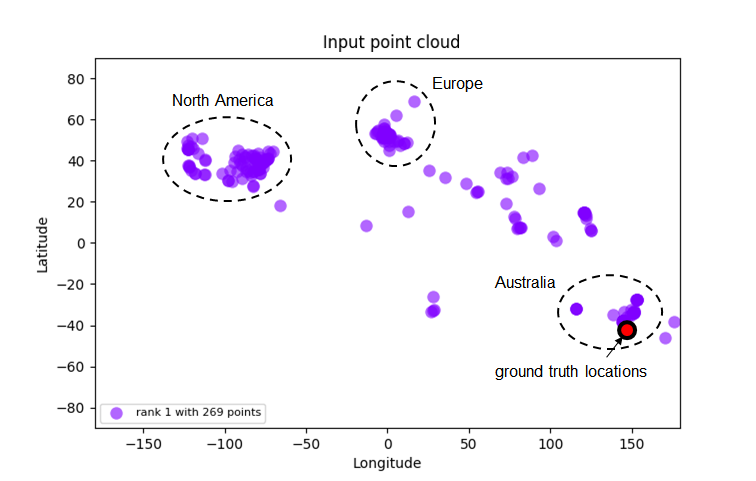}
	\caption{An example input point cloud of ambiguous gazetteer entry locations of a set of place names from the campus dataset, with ground truth locations highlighted in red color.}
	\label{fig:input}
\end{figure}

\subsection{Experiment setup}
A total of 16 algorithms are evaluated based on their performance using the datasets: overall minimum distance (OMD), centroid, minimum distance to unambiguous referents (DTUR), DBSCAN, DBSCAN with automatically determined parameter (\textit{k}-dist), OPTICS, OUTCLUST, CURE, CHAMELEON, HDBSCAN, KMeans, GMM, SNN, Spectral, SOM and DensityK. For \textit{k}-dist, the author did not give a straightforward way to determine a threshold. Therefore, we use the 2$\sigma$ rule in the same way as it is used in DensityK (Algorithm~\ref{alg:1}), to enable a fair comparison. For algorithms that have not been used for place name disambiguation before (i.e., from \textit{k}-dist to SOM), Algorithm~\ref{alg:2} is used on the generated clusters for disambiguation. In case a top-cluster of a place name contains more than one gazetteer entries of this place name, the place name cannot be disambiguated and the case will be regarded as a failure. Different parameters of the algorithms are tested, as shown in Table~\ref{table:parameters}.

\begin{table}[h]
	\centering
	\caption{Parameter configurations of algorithms to be tested for place name disambiguation.}
	\label{table:parameters}
	\resizebox{\textwidth}{!}{%
	\begin{tabular}{l c c c}
		\toprule
		Parameter & Notion & Value & Algorithms \\
		\midrule
		Distance threshold (meters)	& $\varepsilon$ & 200, 2000, 20000  &　DBSCAN \\
		No. of nearest neighbors 	& $k$ & 5, 10, 25 & OUTCLUST, SNN, Chameleon, Spectral\\
		No. of clusters to derive 	& $c$ & 3, 5, 10, 20 & OPTICS, CURE, KMeans, GMM, Spectral\\
		Minimum points in cluster	& $MinPts$ & 1, 5, 10 & DBSCAN, \textit{k}-dist\\
		Frequency threshold 		& $f$ & 0.1, 0.2, 0.5 & OUTCLUST\\
		Weighting coefficient 		& $\alpha$ & 0.1, 1, 10 & Chameleon\\
		SOM dimension          		& $m, n$ & (5, 5), (10, 10), (20, 20) & SOM \\
		\bottomrule
	\end{tabular}}
\end{table}

There is a number of algorithmic features that are important in the place name disambiguation task. The first one is robustness: that an algorithm should ideally work on different input datasets and have mimimum variance in precision and distance error. The next feature is minimum parameter-dependency. A parameter-free algorithm, or an algorithm with parameters automatically determinable, is desirable. Again, this is because for place name disambiguation, no pre-knowledge such as distances between places, or the extent of the space should be assumed for an input. Lastly, an algorithm should also ideally be parameter-insensitive, i.e., modifying parameter values will not lead to significantly different results. Regarding these features, the degree of satisfaction of each of these algorithms when used for fine-grained place name disambiguation will be discussed.

\section{Clustering algorithm performance results} \label{sec:discussion}
Table~\ref{table:result} presents the precision of each algorithm on the tested datasets, and the precisions are based on the best-performing parameter configurations of these algorithms. DensityK achieves the highest precisions, followed by DBSCAN. This is not surprising, as DensityK is designed to be more flexible in determining cluster distances compared to DBSCAN. In the remaining part of this section, the clustering results by each algorithm are discussed individually and compared with each other. This comparison provides a better insight of whether each of these algorithms is suitable for the task of this research, regarding both the feature requirements and performance.

\begin{table}[h]
	\centering
	\caption{Average precision of each algorithm with the best-performing parameters on the tested datasets.}
	\label{table:result}
	\begin{tabular}{l l c}
		\toprule
		Category & Algorithm & Precision \\ 
		\midrule
		\multirow{3}{*}{Ad-hoc} 						& OMD & 76.7\% \\
														& Centroid & 57.2\% \\
														& DTUR & 69.3\% \\
		\midrule
		\multirow{4}{*}{Density-based} 					& DBSCAN & 81.5\% \\
														& DBSCAN \textit{k}-dist & 75.4\% \\
														& OPTICS & 73.2\% \\
														& OUTCLUST & 70.6\% \\
		\midrule
		\multirow{3}{*}{Hierarchical-based} 			& CURE & 78.9\% \\
														& CHAMELEON & 58.3\% \\
														& HDBSCAN & 75.7\%\\
		\midrule
		\multirow{2}{*}{Partitioning relocation-based} 	& KMeans  & 73.4\% \\
														& GMM  & 80.8\% \\
		\midrule
		\multirow{3}{*}{Others} 						& SNN 	& 70.5\% \\	
														& Spectral 	& 74.4\% \\	
														& SOM 	& 73.1\% \\	
		\midrule
		\textbf{The new algorithm}								& \textbf{DensityK} 	& \textbf{83.1\%} \\
		\bottomrule
	\end{tabular}
\end{table}

\begin{figure}[ht]
	\centering
	\includegraphics[width=0.9\textwidth]{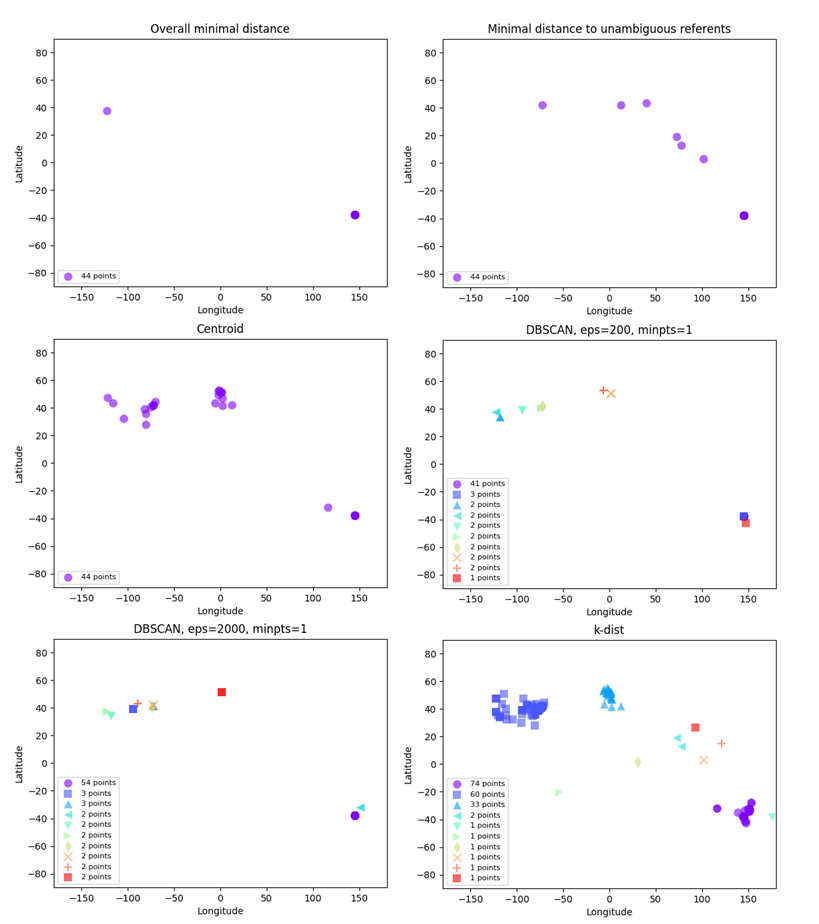}
	\caption{Clustering results generated by established clustering algorithms for place name disambiguation.}
	\label{fig:2.1}
\end{figure}

The clustering results generated by algorithms used for place name disambiguation in the literature, i.e., overall minimum distance, centroid, minimum distance to unambiguous referents, and DBSCAN, are shown in Figure~\ref{fig:2.1}, ranked by number of points contained. A major drawback of the overall minimum distance as well as the minimum distance to unambiguous referents methods is that they are sensitive to \textit{noise place names}: place names with their actual location not captured by gazetteer. For example, the place name \textit{Union House} is referring to a building in the University of Melbourne campus. Its true location has no corresponding gazetteer entry, and the ambiguous gazetteer entries retrieved for this place name in the input point cloud are elsewhere around the world with the same name. Such cases are common for fine-grained place names, while prominent place names (e.g., natural or political) are less likely to be missing in a gazetteer. Another disadvantage of the overall minimum distance method is scalability, as its time cost is significantly larger (over ten times) than other algorithms for most of the dataset tested, particularly for documents with large number of place names and high ambiguities. The centroid-based method performs badly as the input point cloud is spread over the earth, and the centroid is somewhere in the middle and far from the actual focus of the groundtruth locations. 

DBSCAN is robust against noise place names, as it can capture the spatial context (the highlighted red region shown in Figure~\ref{fig:input}) of the original description and neglect entries outside of it. For the example point cloud, when the parameter $\varepsilon$ is set to 2000$m$, the resulting disambiguation precision is higher than with other values selected from Table~\ref{table:parameters}. More groundtruth entries are missed by the cluster generated with a value of 200$m$, and more ambiguous entries are included with a value of 20000$m$. For the clusters generated by the \textit{k}-dist method, the value of $\varepsilon$ determined in this case is roughly 300$km$, which is significantly larger than the most suitable value (somewhere between 1000 and 2000$m$). Consequently, \textit{k}-dist performs badly in this case.
 
\begin{figure}[ht]
	\centering
	\includegraphics[width=0.9\textwidth]{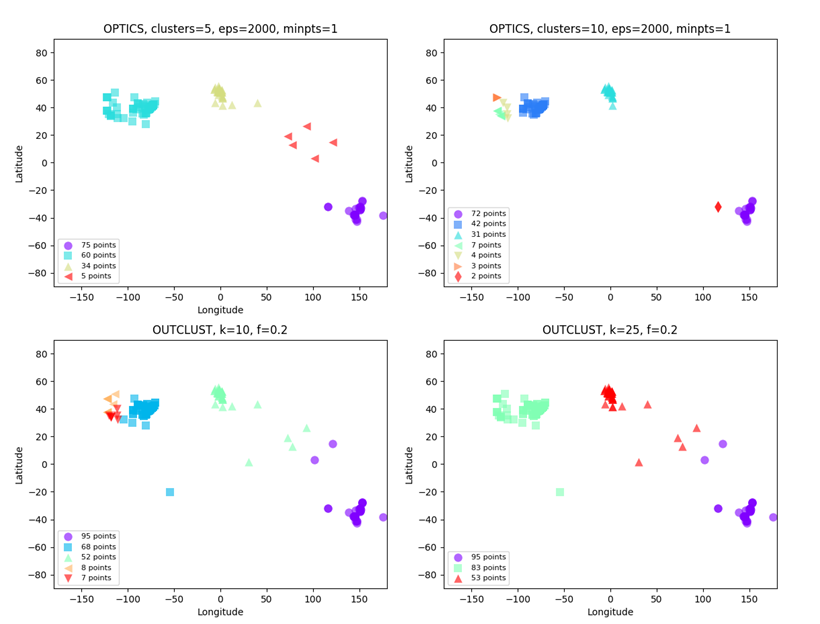}
	\caption{Clustering results generated by density-based clustering algorithms.}
	\label{fig:2.2}
\end{figure}

Figure~\ref{fig:2.2} shows clustering results generated by two other density-based clustering algorithms OPTICS and OUTCLUST for the example input data. OPTICS is designed to overcome the limitation of parameter-dependency of DBSCAN, thus it is expected to perform similar to DBSCAN with the best-performing parameters. The result shows that although OPTICS is more flexible in deriving clusters of various densities based on the tested datasets, this is actually a disadvantage for the task of this research. OPTICS tends to aggregate points from the ground truth spatial context with other points that are relatively close to it, despite that these marginal points have relatively larger local densities. In addition, the parameter \textit{NumberOfClusters} ($c$) of OPTICS is problematic to define. Nevertheless, it is found that setting the value to 10 generally leads to optimal results regardless of input. OUTCLUST has the same drawback of merging nearby points from the spatial context, and it is decided by both parameters $k$ and $f$. The two parameters are more sensitive to input data compared to $c$ of OPTICS, and there is no straightforward method to determine the values either. A large input $k$ value will result in few clusters, as more data points will be regarded as neighbors, and vice versa. Compared to OPTICS, OUTCLUST focuses more on relative density by considering nearest neighbors rather than absolute density, thus, boundary points that are relatively close to some clusters while isolated from others, are more likely to be merged.

\begin{figure}[h]
	\centering
	\includegraphics[width=0.9\textwidth]{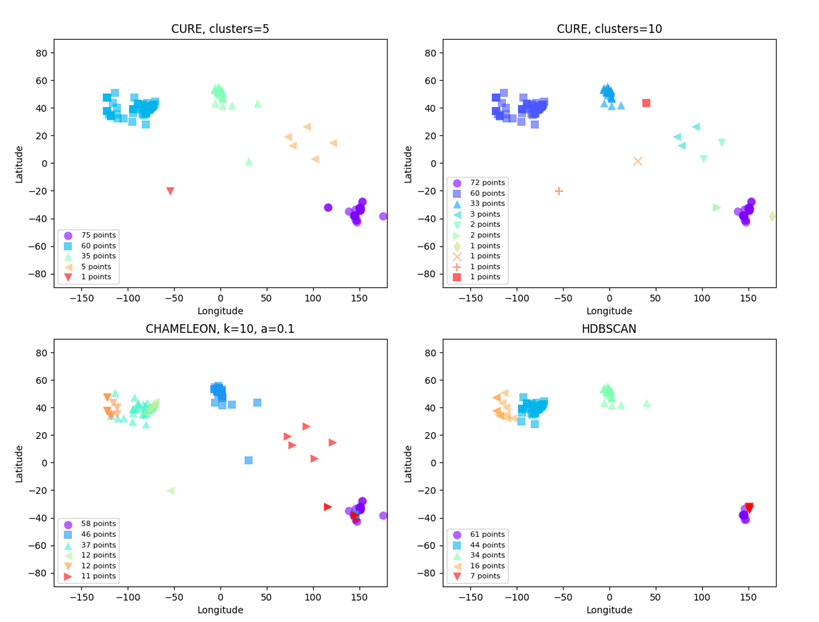}
	\caption{Clustering results generated by hierarchical clustering algorithms.}
	\label{fig:2.3}
\end{figure}

Clustering results by hierarchical clustering algorithms are shown in Figure~\ref{fig:2.3}. CURE requires parameter $c$, similar to OPTICS. The derived clusters by CURE are generally similar to OPTICS. CHAMELEON is more parameter-sensitive than CURE, and the resulting disambiguation precision is not as good as CURE even with the best-performing parameters. As for HDBSCAN, although it does not require any mandatory input parameters, the resulting precision for some input data is only slightly worse than DensityK. However, HDBSCAN is not robust against different input data -- it performs quite well for some data, but significantly worse for others. We discuss this in more detail later in this section.

\begin{figure}[!h]
	\centering
	\includegraphics[width=0.9\textwidth]{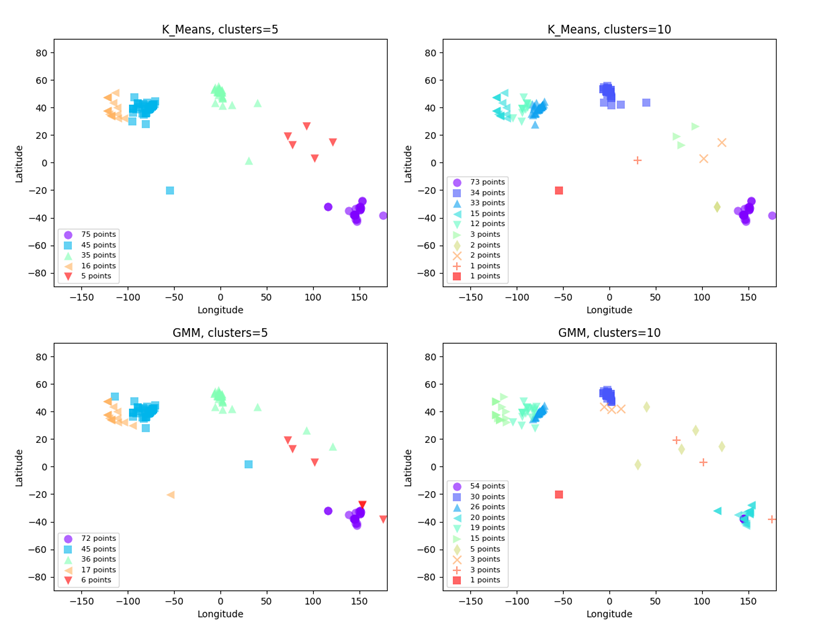}
	\caption{Clustering results generated by partitioning relocation clustering algorithms.}
	\label{fig:2.4}
\end{figure}

Clustering results for using the partitioning relocation-based algorithm are shown in Figure~\ref{fig:2.4}. The KMeans algorithm aims at minimizing inter-cluster distances and dividing the data into $k$ clusters. As a partition-based algorithm, it is not expected to perform well on fine-grained place name disambiguation, which is not a classification problem, and the resulting average precision is worse than HDBSCAN and CURE. For some input data, GMM performs well and achieves the same precisions as DensityK, or as DBSCAN with the best performing parameter values. The performance is generally good (measured by average precision) and robust (e.g., compared to HDBSCAN, which is discussed later). In addition, for most input data, setting different values of $c$, once larger than 10, makes little difference to the clustering compared to algorithms such as KMeans or CURE. Still, there is no easy way to automatically determine the value of $c$, and a single value does not always lead to the highest precisions for different input data. 

\begin{figure}[h]
	\centering
	\includegraphics[width=0.9\textwidth]{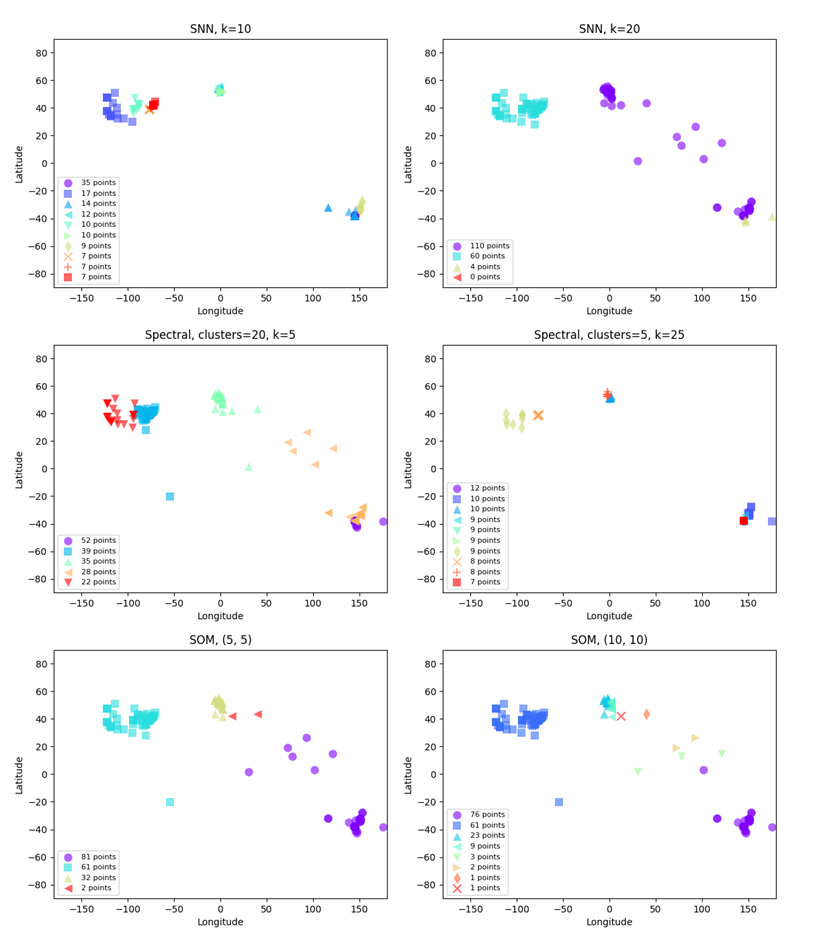}
	\caption{Clustering results generated by other clustering algorithms.}
	\label{fig:2.5}
\end{figure}

Figure~\ref{fig:2.5} shows the results using the remaining three algorithms. SNN is highly sensitive to the parameter $k$, the number of nearest neighbors to consider, and different $k$ values often result in significantly different clustering results, as shown in the figure. A large $k$ value tends to result in only a few large, well-separated clusters, and small local variations in density have little impact. Similar to OUTCLUST, there is no easy way to determine a suitable, meaningful number of nearest neighbors to consider. Spectral clustering also has the problem of parameter sensitivity, both for $c$ and $k$. Its precision is almost always worse than algorithms such as DBSCAN, CURE, and GMM, even with the best-performing parameter values. The resulting clusters generated by SOM are often similar in pattern to those derived by CURE or KMeans, but the average precision is much lower (even lower than Spectral clustering). One advantage of SOM is that the SOM dimension can easily be set to large numbers, which typically leads to higher precisions compared to adopting small values such as $(5, 5)$. When it is set to more than $(20, 20)$, continually increasing the values makes minimal difference to the resulting clusters, as well as precisions.

\begin{figure}[h]
	\centering
	\includegraphics[width=0.6\textwidth]{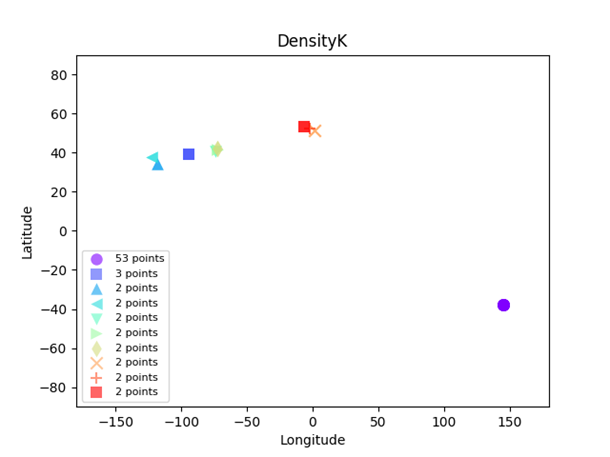}
	\caption{Clustering results generated by the DensityK algorithm.}
	\label{fig:3}
\end{figure}

The result by DensityK is shown in Figure~\ref{fig:3}. The clusters generated are similar to DBSCAN with $\varepsilon$ set to 2000$m$ for this particular input, as shown in Figure~\ref{fig:2.1}. Compared to the results generated by the other algorithms, as shown in Figure~\ref{fig:2.2}, \ref{fig:2.3}, \ref{fig:2.4} and \ref{fig:2.5}, it can be seen that the first-ranking cluster (the purple circles) generated by DensityK is most focused and similar to the highlighted ground truth spatial context shown in Figure~\ref{fig:input}.

\begin{figure}[h]
	\centering
	\includegraphics[width=\textwidth]{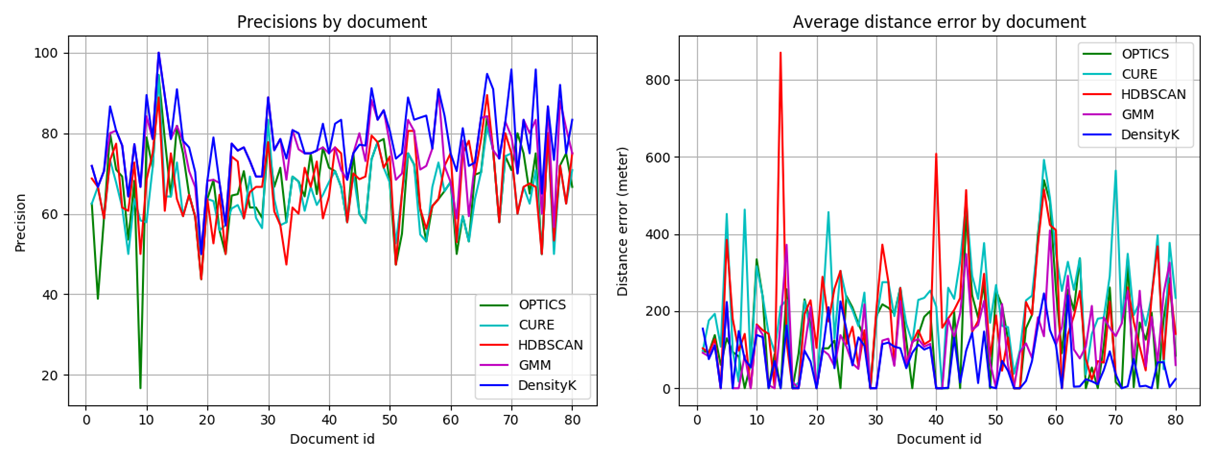}
	\caption{Precision (left), and average distance error in $km$ (right) by description documents.}
	\label{fig:robust}
\end{figure}

From the tested algorithms, OPTICS, CURE, HDBSCAN, GMM, and DensityK seem to be most suitable for place name disambiguation considering the feature requirements. They provide good disambiguation precision, and either do not require input parameters (HDBSCAN and DensityK), or have parameters easy to determine and work well on various input data ($c$ for OPTICS, CURE, and GMM). In comparison, parameters such as \textit{k} or $\varepsilon$ are more sensitive to input, and cannot be determined easily each time a new input is given. Here we further evaluate the robustness of the five algorithms over different input data, in terms of variation in precision and average distance error, i.e., the average distance between the ground truth locations of place names and the entries selected by these algorithms. We randomly select documents from our dataset, and the results are shown in Figure~\ref{fig:robust}. DensityK has almost always the highest precision, as well as low variation compared to the other algorithms, particularly HDBSCAN and OPTICS. In terms of distance errors, DensityK has the least variance as well as overall minimum distance errors.

\begin{figure}[h]
	\centering
	\includegraphics[width=\textwidth]{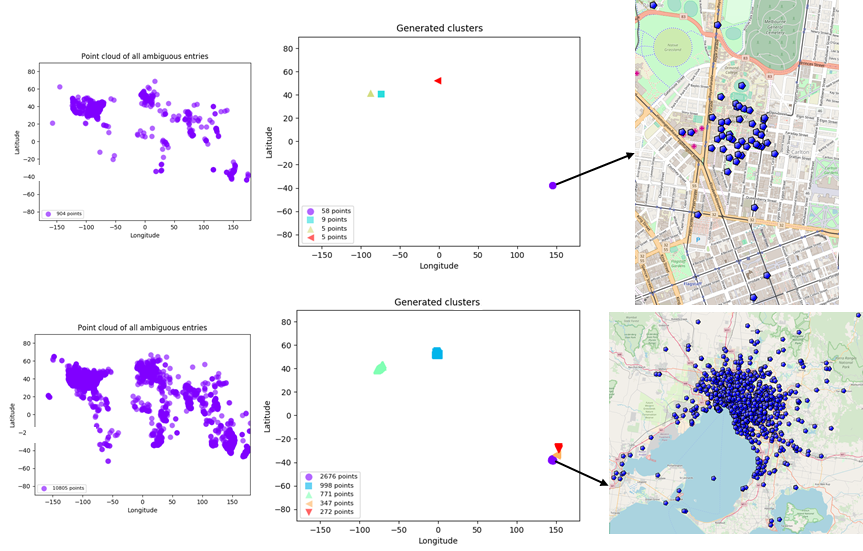}
	\caption{Clusters derived for merged places from the campus dataset (top) and the Melbourne dataset (bottom) representing spatial contexts. The left hand side shows the input point clouds of all ambiguous entries.}
	\label{fig:CR}
\end{figure}

Figure~\ref{fig:CR} shows the clustering results for places merged from the two dataset using DensityK, representing the spatial contexts of the two data sources where the descriptions are embedded, i.e., the University of Melbourne campus, and Melbourne.

\section{Conclusions} \label{sec:conclusion}
Place descriptions in everyday communication provide a rich source of knowledge about places. In order to utilize such knowledge in information systems, an important step is to locate the places being referred to. The problem of locating place names from text sources is often called toponym resolution, which consists of two tasks: place name identification from text, and place name disambiguation. This research looks at the second task, and more specifically, disambiguating fine-grained place names extracted from place descriptions. We focus on clustering-based disambiguation approaches, as clustering approaches require minimum pre-knowledge of the place names to be disambiguated compared to knowledge- and machine learning-based approaches. 

For this purpose, we first select clustering algorithms that have been used for place name disambiguation in the literature, or are from other communities (e.g., data mining) and are regarded as promising for this task. We evaluate and compare the performance of these algorithms based on two different datasets using precision and distance error. For algorithms that require parameters, different values of each parameter are tested in a grid-search manner. We then analyze the performance and associated causes for each algorithm, its parameter-dependency and parameter-sensitivity, robustness (in terms of variance of their performance over different input data), and discuss the suitability of each algorithm for fine-grained place name disambiguation based on these criteria.

Furthermore, a new clustering algorithm, DensityK, is presented. It is designed to overcome several identified limitations of the previous algorithms. It out-performs the other tested algorithms and achieves state-of-art disambiguation precision on the test datasets. The algorithm is based on analyzing local densities of an input point cloud, which consists of all ambiguous gazetteer entries corresponding to the place names extracted from an input document. It then derives a density threshold for determining clusters that have significantly larger densities than other points. Compared to the other algorithms, DensityK is parameter-independent, robust against different input data with various spatial extents, densities, and granularities, which makes it most desirable for the task of this research. This is reflected by consistently achieving higher precision and overall minimum distance error compared to other competitive algorithms. The worst time complexity of the algorithm is same as DBSCAN ($O(n^2)$), when both are considered without any indexing mechanism for neighborhood queries. The time complexity is better than algorithms such as overall minimum distance clustering.

The focus of this research is to provide recommendations for the selection of appropriate methods of clustering-based disambiguation, for fine-grained place names from place descriptions. We have not yet considered further optimizing the developed algorithm, although we explained briefly how indexing and optional parameters can be used for facilitating processing time in Section~\ref{3.2}. Optimization is important considering certain applications such as processing streaming data for goals such as geographic information retrieval. Finally, a clustering algorithm for this purpose can be used in conjunction with other knowledge- or machine-learning based approaches to enhance precision, which is beyond the scope of this research.

\bibliographystyle{spbasic}
\bibliography{bibfile}
\end{document}